\documentclass[twocolumn]{aastex63}
\usepackage{graphicx}
\usepackage{subfigure}
\usepackage{color, hyperref, epsfig}
\usepackage{apjfonts, natbib}
\usepackage{appendix}
\usepackage{float}
\usepackage{bm}
\usepackage{tabularx}
\usepackage{makecell}
\usepackage{multirow}
\usepackage{lineno}

\newcommand{\lum}{erg~s\ensuremath{^{-1}}}

\newcommand{\rosat}{\emph{ROSAT}}

\newcommand{\flux}{erg~s$^{-1}$~cm$^{-2}$}

\newcommand{\msun}{\ensuremath{\rm M_{\odot}}}
\newcommand{\myr}{$\rm M_\odot\ yr^{-1}$}

\newcommand{\kms}{\ensuremath{\mathrm{km~s^{-1}}}}
\newcommand{\mbh}{\ensuremath{M_\mathrm{BH}}}

\newcommand{\wise}{\emph{\it WISE}}
\newcommand{\neowise}{NEOWISE}

\newcommand{\vdisp}{\ensuremath{\sigma\mathrm{_{\star}}}}

\newcommand{\objectfullname}{SDSS J010320.39+140152.5}
\newcommand{\objectshortname}{J0103+1401}
\shorttitle{J0103}
\shortauthors{Liu et al.}
\newcommand{\affone}{Department of Astronomy, University of Science and Technology of China,
Hefei, 230026, China; wdgj@mail.ustc.edu.cn,jnac@ustc.edu.cn}
\newcommand{\afftwo}{School of Astronomy and Space Sciences,
University of Science and Technology of China, Hefei, 230026, China}
\newcommand{\affthree}{Department of Physics, Anhui Normal University, Wuhu, Anhui, 241002, China; sunluming@ahnu.edu.cn}
\newcommand{\afffour}{IPAC/Caltech, 1200 E. California Boulevard, Pasadena, CA 91125, USA}
\newcommand{\afffive}{Department of Astronomy, Guangzhou University, Guangzhou 510006, China}

\submitjournal{\apj}
\received{Feb 2, 2026}
\accepted{May 26, 2026}

\begin{document}

\title{An Obscured Tidal Disruption Event Uncovered by Its Mid- and Near-Infrared Dust Echo in a Star-Forming Galaxy}
\author[0000-0002-1438-2199]{Hui Liu}
\affiliation{\affone}
\affiliation{\afftwo}

\author[0000-0002-7223-5840]{Luming Sun}
\affiliation{\affthree}

\author[0000-0002-7152-3621]{Ning Jiang}
\affiliation{\affone}
\affiliation{\afftwo}

\author[0000-0002-7020-4290]{Xinwen Shu}
\affiliation{\affthree}

\author[0000-0003-4225-5442]{Yibo~Wang}
\affiliation{\affone}
\affiliation{\afftwo}

\author[0000-0002-1517-6792]{Tinggui Wang}
\affiliation{\affone}
\affiliation{\afftwo}

\author[0000-0002-0077-2305]{Roc M. Cutri}
\affiliation{\afffour}

\author[0000-0002-4757-8622]{Liming Dou}
\affiliation{\afffive}

\author[0009-0003-2754-6898]{Fabao Zhang}
\affiliation{\affthree}

\author[0000-0003-3824-9496]{Jiazheng Zhu}
\affiliation{\affone}
\affiliation{\afftwo}

\author[0000-0001-6938-8670]{Zhenfeng Sheng}
\affiliation{Institute of Deep Space Sciences, Deep Space Exploration Laboratory, Hefei, 230026, People’s Republic of China}


\begin{abstract}
We present a comprehensive study of an infrared (IR) flare in the star-forming galaxy SDSS J010320.39+140152.5, which is selected from the sample of mid-IR (MIR) outbursts in nearby galaxies (MIRONG). Its MIR luminosity rose rapidly to a peak of $\sim5.4\times10^{43}$ \lum, maintained in the high state for about a year, and decreased continuously afterward. No optical variability was detected throughout the IR flare. Near-IR follow-up observations around the peak pinpointed the flare's location to spatially coincide with the galactic nucleus, with a $3\sigma$ upper limit of the offset of $\lesssim100$ pc. The IR spectral energy distribution (SED) of the flare is consistent with thermal emission of dust with temperatures of $\sim900$ K.
Using a dust radiative transfer model, we inferred a peak UV luminosity of $\sim(4-10)\times10^{44}$ erg s$^{-1}$ and a total energy of $\sim(0.9-2)\times10^{52}$ ergs released.
We ruled out the possibility of a supernova, and prefer that the IR flare originated from an obscured tidal disruption event (TDE) rather than a changing-look active galactic nucleus (AGN). This flare stands as one of the most compelling cases to date for the emerging class of dust-obscured TDEs in recent years. They are missed by optical surveys, partly accounting for the observed bias in TDE host galaxies, and represent a crucial, yet often overlooked, component for a complete understanding of the TDE population.
\end{abstract}


\section{introduction}
Tidal disruption events (TDEs) occur when stars come too close to supermassive black holes (SMBHs) and are tidally disrupted \cite{Hills1975}.
About half of the stellar debris falls back to the central SMBH, producing a flare that peaks at the UV to soft X-ray band (\citealt{Rees1988, Phinney1989}).
The first TDE was detected in the X-ray band (\citealt{Bade1996}) with the \rosat\ telescope, while optical time-domain surveys have dominated the discovery of TDEs (\citealt{vV2020, vV2021, Gezari2021}) in recent years.
The total number of TDEs has exceeded $\sim200$.

Statistically, optical TDEs display a puzzling distribution across different types of host galaxies: they are overrepresented in post-starburst galaxies (\citealt{Arcavi2014, French2016, French2020}), which experience a violent starburst episode approximately $10^9$ yrs previously (\citealt{Dressler1983}), but are almost absent in normal star-forming galaxies.
Scenarios that may lead to the rate enhancement have been proposed (see the review in \citealt{French2020}), such as SMBH binaries (\citealt{Chen2009, Coughlin2019}), central stellar over-density (\citealt{French2020, Stone2016}), and velocity anisotropy (\citealt{Stone2018}). However, these explanations cannot address why TDEs are rare in galaxies with current intense star formation.

Star-forming galaxies are long known to have a large amount of dust, which may obscure the TDE occurring in their nuclei.
Fortunately, the nuclear dust can reprocess the UV radiation it absorbs into IR reradiation, hence forming an echo.
Such IR echoes have been observed \cite{Jiang2016,vV2016, Dou2016} by the IR time-domain survey Wide-field Infrared Survey Explorer (\wise; \citealt{Wright2010, Mainzer2014}).
IR echo provides a new probe to detect obscured TDEs, and the first case is Arp 299-B AT1 \citep{Mattila2018}, whose TDE nature was confirmed with the help of a transient radio jet.
Subsequent studies have identified dozens of TDE candidates in star-forming and starburst galaxies using IR surveys (e.g., \citealt{Kool2020, Jiang2021a, Reynolds2022, Masterson2024}).
Unfortunately, the nature of the vast majority of these TDE candidates still lacks clear validation.

Most IR flares lack optical counterparts \citep{Jiang2021a, Masterson2024}.
Thus, the TDE rate in star-forming galaxies would be greatly underestimated if the majority of IR flares are indeed TDEs.
This suggests that the rare optical detection of TDEs in star-forming galaxies could be a selection bias due to dust obscuration.
However, to prove this point, it is necessary to verify the TDE nature of these IR flares.
Moreover, a tool that can infer the UV peak luminosity based on the IR data is required to compare the luminosity functions of optical TDEs and IR TDEs.


In this paper, we present the multi-band analysis of an IR TDE candidate in a nearby spiral star-forming galaxy \objectfullname\ (RA = 01:03:20.39, DEC = +14:01:52.5, hereafter \objectshortname) at $z=0.04181$.
The transient was first reported as an IR transient in the sample of MIR Outbursts in Nearby Galaxies (MIRONG; \citealt{Jiang2021a}), and was not alerted by any optical survey.
We describe the public data and our follow-up observations in Section~\ref{Sec 2}, present the analysis of multi-bands photometric and spectroscopic data in Section~\ref{Sec 3}, and describe a dust echo model in Section~\ref{Sec 4}, and discuss the nature of the IR transient and the implications in Section~\ref{Sec 5}, and finally summarize in Section~\ref{Sec 6}.
Throughout this paper, we assumed a $\rm \Lambda CDM$ cosmology with $\Omega_{\rm M} = 0.3$, $\Omega_{\Lambda} = 0.7$, and $H_{0} =70$ km~s$^{-1}$~Mpc$^{-1}$.

\section{Data and observations} \label{Sec 2}

\subsection{MIR discovery} \label{Sec 2 mir discovery}

\objectshortname\ was discovered using data from the all-sky survey \wise\ and its successor Near-Earth Object Wide-field Infrared Survey Explorer  (\neowise).
The original \wise\ survey has mapped the full sky every $\sim$ 180 days, working in 4 bands (W1, W2, W3, W4) centered at 3.4, 4.6, 12, 22 $\mu$m since Jan 2010 until its hydrogen cryostat was exhausted.
The survey was extended by an additional 4 months as \neowise\ Post-Cryogenic Mission operated only in W1 and W2 bands.
Following a 33-month hibernation period, the \wise\ instrument recommenced survey operation in Dec 2013, and this post-hibernation mission is referred to as \neowise\ Reaction.
The \wise\ and \neowise\ surveys visit the object about half a year in W1 and W2 with $\sim 10-20$ exposures per band per visit.

\begin{figure*}[!t]
\centering
\begin{minipage}{1.0\textwidth}
\centering{\includegraphics[angle=0,width=1.0\textwidth]{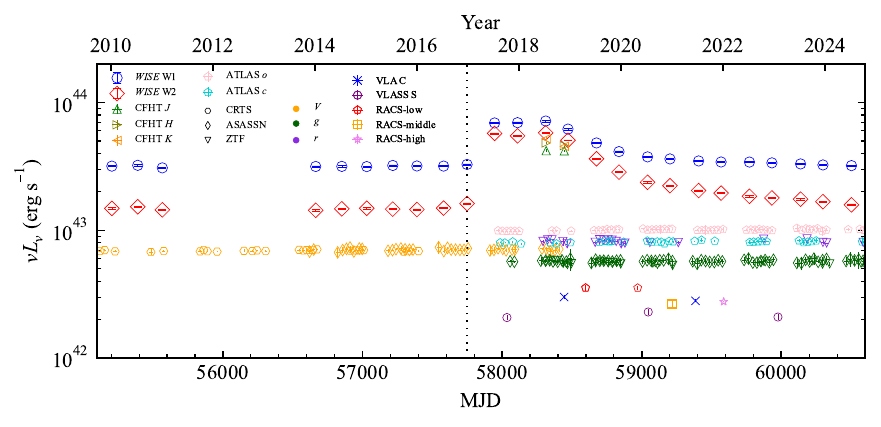}}
\end{minipage}
  \caption{The multi-wavelength light curves of \objectshortname\ in \wise\ W1/W2 (blue circles/red diamonds), CFHT $J$/$H$/$K$ (green up/olive right/orange left triangles), ATLAS $o/c$ (brown/cyan pentagons), CRTS $V$ (orange hexagons), ASASSN $V$/$g$ (orange/green thin diamonds), ZTF $g$/$r$ (green/violet down triangles), VLA C (blue x, VLASS S (purple circles), and RACS-low (red pentagons), middle (orange squares), high (magenta stars) bands.
  We added a constant magnitude offset of $+0.574$ for CRTS to match the ASASSN $V$-band median magnitude.
  For clarity, we divided the optical luminosities by a factor of 10, and multiplied the radio ones by a factor of $10^4$.
  The vertical dotted line indicates the MIR detection of the flare.
  }
\label{fig_lightcurve}
\end{figure*}

We adopted unWISE coadds (\citealt{Lang2014, Meisner2018}) that stack all single exposures during a specific visit to improve the signal-to-noise ratio (S/N).
The unWISE data up to 2024 has been released, with a total of 25 visits, which we referred to as epoch 1 to 25, respectively.
We measured the photometry of the whole galaxy from an isophotal $\rm K_S$ fiducial aperture ($r\_k20fe$) obtained in the 2MASS all-sky source catalog (the same aperture will be applied in other bands when carrying out the host galaxy fluxes with no further explanation).
We show the MIR light curves in the W1 and W2 bands in Figure~\ref{fig_lightcurve}.
The galaxy remained invariant until an abrupt rise in 2017.

\subsection{Archival optical light curve data}

We collected archived optical light curve data of \objectshortname\ as follows.

The ongoing All-Sky Automated Survey for Supernovae (ASASSN; \citealt{Shappee2014}) has been observing the galaxy since 2013.
We obtained image subtraction photometry (reference flux added) in the $V$ and $g$ bands from ASASSN Sky Patrol (\citealt{Kochanek2017}), and binned the light curves monthly with a few to 20 data points per bin for a deeper sensitivity.
As shown in Figure~\ref{fig_lightcurve}, no significant variation was detected around 2017, when the IR transient occurred.

To check for possible optical flares over a longer period of time, we collected data from the Catalina Real-time Transient Survey (CRTS; \citealt{Drake2009}), the Zwicky Transient Facility (ZTF; \citealt{Masci2018}), and the Asteroid Terrestrial-impact Last Alert System (ATLAS; \citealt{Tonry2018}).
The CRTS data from 2005 to 2014 are public, and the ongoing ZTF and ATLAS surveys provide data starting in $\sim2018$.
We obtained the CRTS $V$-band light curve from CRTS data release 3.  
We downloaded the $g$- and $r$-band science images of ZTF from the Infrared Science Archive (IRSA), and ATLAS $c-$ and $o-$band science images from NASA Planetary Data System (PDS), and measured the aperture flux. We show the CRTS, ZTF, and ATLAS light curves in Figure~\ref{fig_lightcurve}, which are also binned monthly.
The galaxy showed no significant optical variability in the past $\sim 20$ years.

\subsection{Follow-up NIR imaging observations}

We triggered follow-up NIR observations of \objectshortname\ on Jul 16 (epoch 1) and Nov 22, 2018 (epoch 2) in the $J$-, $H$-, and $K$-band with the Wide-field InfraRed Camera (WIRCam) on the Canada France Hawai'i Telescope (CFHT).
For each epoch, we took three exposures per band, with an exposure time of a single exposure of 30, 15, and 11 seconds for the three bands, respectively.
We computed astrometric solutions using a combination of \texttt{SExtractor} (\citealt{Bertin1996}) and \texttt{SCAMP} (\citealt{Bertin2006}).
We then resampled and combined the images to improve the S/N for each band in each epoch using \texttt{SWarp} (\citealt{Bertin2002}).

To remove the host galaxy contribution and obtain the flux of the flare, we collected NIR images taken in August 2007, before the IR flare occurred, from the UKIRT Infrared Deep Sky Survey (UKIDSS; \citealt{Lawrence2007}) as reference images.
We downloaded the images from the WFCAM Science Archive.


\subsection{Spectroscopic Observations}

We obtained a historical optical spectrum of the galaxy observed by Slone Digital Sky Survey (SDSS), which was taken on Sept 24, 2000, before the IR flare occurred.

After the IR flare occurred, we took two new optical spectra with the Double Spectrograph (DBSP) mounted on Hale 5m (P200) telescope at Palomar Observatory on Dec 16, 2017 and Sept 2, 2021.
The first was taken around the peak of the MIR flare, and the second was taken during the fading stage, with an exposure time of 900 and 1600 seconds, respectively.
For both spectra, we adopted a D55 dichroic, a blue grating of 600 lines per mm blazed at 3780~\AA, and a red grating of 316 lines per mm blazed at 7150~\AA, resulting in a continuous wavelength coverage of $3100 - 10500$~\AA.
We used slit widths of $1.^{\prime\prime}5$ and $2.^{\prime\prime}0$, respectively, based on the weather conditions.
The spectral resolutions are 7.84/10.46~\AA\ in the red channel and 4.04/5.32~\AA\ in the blue channel.
We reduced the DBSP data according to the standard reduction procedures for a long-slit spectrum.
The flux calibration was made using a standard star's spectrum taken on the same night.

\subsection{X-ray Observations}

To check for any X-ray emission associated with the MIR flare, we triggered two $Swift$ (\citealt{Burrows2005}) observations on Jun 5, 2019 and Dec 16, 2021 with net exposure times of 1666 and 4031 seconds, respectively.
We reprocessed the XRT event files in Photon Counting mode with the task \texttt{xrtpipeline}.
We extracted source photons from a circle centered at the optical nucleus with a radius of $36^{\prime\prime}$ to avoid covering a companion galaxy, and extracted background photons from a source-free annulus region with inner and outer radii of 40 and 60 pixels, respectively.

Neither observation yielded a significant detection ($\lesssim 1\sigma$).
We stacked two observations, and obtained a $3\sigma$ upper limit of the net count rate in the $0.3-10\rm\ keV$ band of $1.7 \times 10^{-3}\ \rm s^{-1}$.
Assuming a power-law spectrum with a photon index of $\Gamma =2$ with a galactic absorption of $\rm N_{H} = 3.55\times10^{20}\ cm^{-2}$ (\citealt{HI4PI2016}), we estimated with \texttt{PIMMS} a $3\sigma$ upper limit for the unabsorbed flux at $0.3-10\rm\ keV$ to be $6\times10^{-14}$ \flux, corresponding to a luminosity of $2.4\times10^{41}$~\lum.

\subsection{Radio Observations} \label{Sec 2 radio observation}

\begin{deluxetable}{lcclcr}
\setlength{\tabcolsep}{0.06in}
\tablecaption{Radio Observations of \objectshortname}
\label{tbl_radio_observation}
\tablewidth{0pt}
\tablehead{
\colhead{Instrument} & \colhead{MJD} &Num.& \colhead{Frequency} &\colhead{\hspace{0.5cm}} & \colhead{Flux} \\
                     &             &  &  \colhead{(MHz)} &    &
\colhead{(mJy)}}
\startdata
               &  58597-58971 & 3 & 887.5 & & 9.81$\pm$0.49 \\
    RACS            & 59216  & 2 & 1367.5        & & 4.77$\pm$0.21 \\
               & 59586 & 2& 1655.5   & & 4.10$\pm$0.05 \\
    NVSS        & 49292  & 1 & 1400      & & $<5.3$ \\
    \multirow{3}{4em}{VLASS}       &  58034    & 1& \multirow{3}{4em}{3000} & & 1.70$\pm$0.08  \\
                  &   59046  & 1 &   &  & 1.88$\pm$0.10  \\
                  &   59975  & 1 &  &  & 1.72$\pm$0.08 \\
    \multirow{2}{4em}{VLA}   & 58444 & 1 & \multirow{2}{4em}{5500} & & 1.34$\pm$0.05 \\
                   &   59385  & 1 &    & & 1.25$\pm$0.03\\
\enddata
\tablecomments{Summary of all radio observations of \objectshortname\ showing where (Instrument) and when (MJD) the observations were done. The columns $``\rm Num."$ and $``\rm Frequency"$ describe the exposure number and the central frequency of the corresponding observation. The last column represents the integrated flux and flux error.}
\end{deluxetable}

We took follow-up radio observations of \objectshortname\ at C-band (central frequency of 5.5 GHz) with the Very Large Array (VLA; \citealt{Perley2011}) in its moderately compact C configuration on Nov 21, 2018 (program code, 18B-086) and Jun 20, 2021 (program code, 21A-146).
The data were reduced following standard procedures with the \texttt{CASA} package.
For both epochs, flux density calibration was conducted using 3C48, whereas the nearby source J0112+2244 was used to determine complex gain solutions that were interpolated to \objectshortname.
After removing the RFI, the data were imaged using the 
\texttt{CLEAN} algorithm, with Briggs weighting and ROBUST parameter of 0.
The final cleaned map suggests a deconvolved source size $2.^{\prime\prime}0\times1.^{\prime\prime}4$ for the first observation and $2.^{\prime\prime}2\times1.^{\prime\prime}4$ for the second observation.
\objectshortname\ is detected as a compact radio source, with an integrated flux density of $1.34\pm0.05$ and $1.25\pm0.03$~mJy, respectively, measured using the \texttt{CASA} task \texttt{IMFIT}.

We obtained public radio data of \objectshortname\ as follows.
The object was observed by the Very Large Array Sky Survey (VLASS; \citealt{Lacy2020}) in the S-band (central frequency of 3 GHz) on Oct 8, 2017, Jul 16, 2020, and Jan 31, 2023.
The Quick Look Stokes $I$ images also show unresolved emission at a resolution of $2.^{\prime\prime}5$.
We measured integrated fluxes of $1.70\pm0.08$, $1.88\pm0.10$ and $1.72\pm0.08$~mJy for the three epochs, respectively.
It was also observed by the Rapid ASKAP Continuum Survey (RACS; \citealt{McConnel2020}) in the RACS-low (887.5 MHz), mid (1367.5 MHz), and high (1655.5 MHz) bands.
During the first survey, \objectshortname\ was scanned 3/2/2 times in each band in Apr 2019, May 2020/Jan 2021/Jan 2022, respectively.
The data generated from the CASDA Data Access Portal show unresolved compact radio cores detected in all observations with mean integrated fluxes of $9.81\pm0.49$, $4.77\pm0.21$, and $4.10\pm0.05$ mJy in each band.

Before the IR flare occurred, \objectshortname\ was observed by the NRAO VLA Sky Survey (NVSS; \citealt{Condon1998}) in 1993.
Due to the poor spatial resolution ($45^{\prime\prime}$), we cannot reliably deblend it from its companion, which shows a comparable level of diffuse radio flux.
Thus, we adopted the pixel flux of $5.3$ mJy as the upper limit of the flux of \objectshortname.
We summarize all radio observations in Table~\ref{tbl_radio_observation}.

\subsection{Host's SED}

\begin{deluxetable}{lccr}
\setlength{\tabcolsep}{0.06in}
\tablecaption{SED of the host galaxy}
\label{tbl_host_sed}
\tablewidth{0pt}
\tablehead{\colhead{Instrument} & Obs Time & \colhead{Filter} & \colhead{Flux (mJy)}}
\startdata
\multirow{2}{4em}{GALEX} & 
\multirow{2}{4em}{2006} 
                         & FUV & $0.178\pm0.013$ \\
                    &    & NUV & $0.281\pm0.011$ \\
SDSS                     & 
\multirow{1}{4em}{1999}
                         & $u$ & 0.84 $\pm$ 0.01 \\
\multirow{5}{4em}{Pan-STARRS} &
\multirow{5}{4em}{2009--2014}   
                         & $g$ & 2.83 $\pm$ 0.04 \\
                    &    & $r$ & 4.50 $\pm$ 0.14 \\
                    &    & $i$ & 6.74 $\pm$ 0.07 \\
                    &    & $z$ & 7.92 $\pm$ 0.15 \\
                    &    & $y$ & 9.16 $\pm$ 0.17 \\
\multirow{4}{4em}{UKIDSS}  &
\multirow{4}{4em}{2007}
                         & $Y$ & 10.87$\pm$0.20\\
                    &    & $J$ & 12.31$\pm$0.13\\
                    &    & $H$ & 15.62$\pm$0.19\\
                    &    & $K$ & 13.74$\pm$0.32\\
\multirow{4}{4em}{$WISE$} &
\multirow{4}{4em}{2010}
                         & W1  &  8.75$\pm$0.26 \\
                    &    & W2  &  5.61$\pm$0.20 \\
                    &    & W3  & 28.94$\pm$0.50 \\
                    &    & W4  & 93.85$\pm$2.50 \\
\enddata
\tablecomments{The SED data before the flare occurred. We show the instruments, the observational time, the filters and the Galactic-extinction-corrected fluxes (errors).}
\end{deluxetable}

\begin{figure*}[htp]
\centering
\begin{minipage}{1.0\textwidth}
\centering{\includegraphics[width=1.0\textwidth]{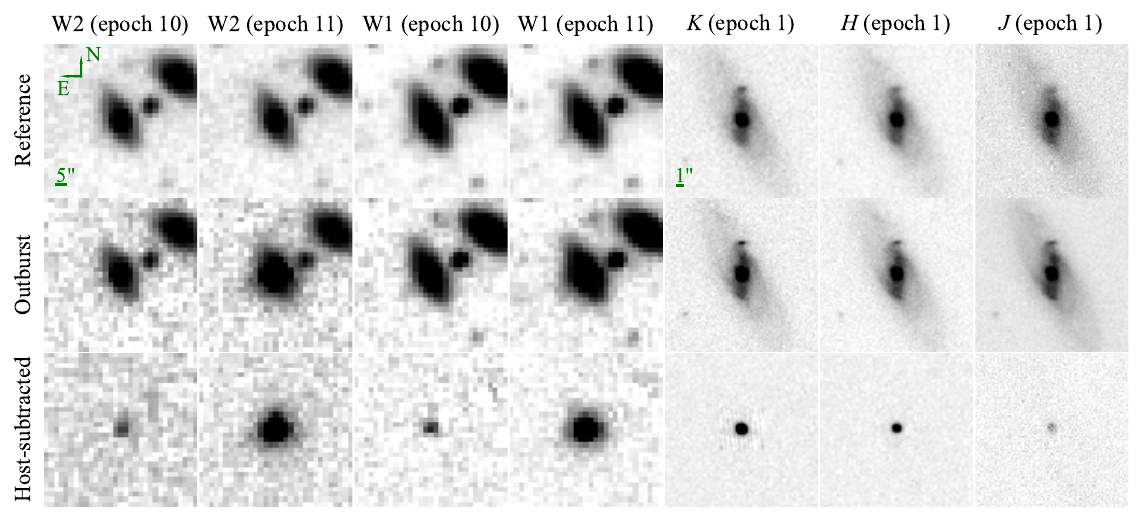}}
\end{minipage}
  \caption{Imaging subtraction results.
  We show results in W2 and W1 bands at epochs 10 (first detection) and 11, and those in $K$, $H$, and $J$ bands at epoch 1.
  From top to bottom rows are the quiescence states, flare states, and the differences between them.
  All the cutouts are centered at the optical center of the host galaxy.}
\label{fig_IRimage}
\end{figure*}

We collected the pre-flare SED data of the host galaxy from UV to MIR from public archives.
For the UV bands, we obtained FUV/NUV fluxes measured by GALEX\footnote{http://galex.stsci.edu/GR6/}.
For the optical bands, we obtained Kron-aperture magnitudes in the $g$, $r$, $i$, $z$, and $y$ filters taken by Pan-STARRS\footnote{https://catalogs.mast.stsci.edu/panstarrs/}, and the model magnitude in the $u$ filter taken by SDSS\footnote{We did not use the SDSS photometry in the $g$, $r$, $i$, and $z$ filters because they agree with the Pan-STARRS photometry while have greater errors.}.
For the NIR bands, we obtained aperture magnitudes in the $Y$, $J$, $H$, and $K$ filters taken by UKIDSS.
And for the MIR bands, we measured aperture photometry in W1 and W2, and obtained profile-fitting magnitudes in W3 and W4 from the ALLWISE catalog.
We list the SED data in Table~\ref{tbl_host_sed}, along with the observational time.
All fluxes have been corrected for Galactic extinction of $\rm E(B-V)=0.048$ in the line of sight of \objectshortname\ (\citealt{Barbary2021, Green2018}).

\section{Data Analysis} \label{Sec 3}

\subsection{Occurrence time and position of the flare}

With the MIR light curves obtained from photometry of the entire galaxy (Figure 1), it is clear that epoch 11 and the subsequent epochs were all in the flare state, while it is not clear whether epoch 10 is in the flare state or not because of the contamination from the host galaxy.
To obtain a more accurate flux of the flare, we made image subtraction for epoch 10 and later epochs, using the median stack of images at epochs from 1 to 9 as the reference images.
The image subtraction was made using \texttt{HOTPANTS} (\citealt{Becker2015}).
We present examples of image subtraction results in Figure~\ref{fig_IRimage}, where a point source is clearly shown on the difference image.

We then performed point spread function (PSF) photometry on the host-subtracted images.
In addition to the statistical error of the photometry, we also evaluated the systematic error caused by the galaxy background in the image subtraction process.
Taking epoch 10 as an example, we performed PSF photometry on difference images of $10-1$, $10-2$, ..., and $10-9$, and adopted the standard deviation of the 9 fluxes as the systematic error.

The flare was significantly detected at epoch 10 ($\gtrsim10\sigma$), and thus we adopted the time of epoch 10, ${\rm MJD}=57748.7$, as the occurrence time of the flare.
In addition, the photometric results were used for the subsequent SED and variation analysis of the flare.

We also made image subtraction on CFHT images in NIR using UKIDSS images as references.
During this process, the $\texttt{kernel}$ matching the effective PSFs (ePSFs) of CFHT and UKIDSS ($\rm ePSF_{ukidss} = ePSF_{cfht} \otimes \texttt{kernel}$) could not be accurately calculated using the default algorithm of \texttt{HOTPANTS}, possibly because the overlap regions of UKIDSS and CFHT images are small and the number of stars for matching the ePSF is not sufficient.
Thus, we manually built ePSFs from nearby stars without restricting the use of overlapping stars, and calculated the $\texttt{kernel}$, both using algorithms in \texttt{Photutils} (\citealt{Bradley2025}).
The flare was detected significantly in the $K$ and $H$ bands with $>25\sigma$, while only barely detected in the $J$ band with a few $\sigma$.

\begin{figure*}[htp]
\centering
\begin{minipage}{1.0\textwidth}
\centering{\includegraphics[angle=0,width=1.0\textwidth]{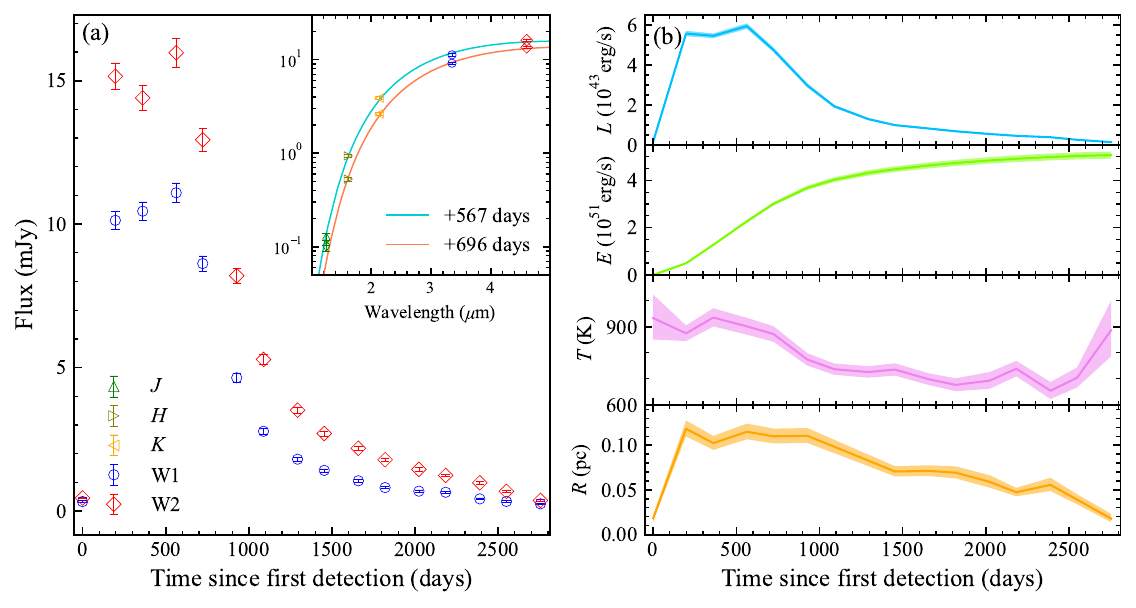}}
\end{minipage}
\caption{SED analysis of the IR flare. \textbf{a,} The MIR light curves and the NIR-to-MIR SEDs of the flare. \textbf{b,} From top to bottom rows are the blackbody luminosity, the integrated energy, the blackbody temperature, and the blackbody radius of the flare.}
\label{wisebbfit}
\end{figure*}

We measured the positions of the flare and the nucleus of the host galaxy by fitting the difference and reference images with a 2-dimensional Gaussian, respectively, and then calculated the offset between them.
For NIR bands, we adopted the positions and offset measured from the $K$ band with the best S/N.
The uncertainty is dominated by the systematic error of the astrometric calibration, and we estimated it from the offsets of nearby stars between UKIDSS and CFHT images after correcting for the stars' proper motions (\citealt{gaia2018}).
The position of the flare spatially coincides with the galaxy nucleus because we measured an offset of $\sim0.^{\prime\prime}041\pm0.^{\prime\prime}038$, corresponding to a projected distance of $37\pm34$ pc in physical.
The flare's position measured in the MIR W2 band also agrees with the NIR result, with an offset of $\sim0.^{\prime\prime}12\pm0.^{\prime\prime}08$.

\subsection{SED of the flare and luminosity} \label{Sec 3 sed of the flare and luminosity}

We analyzed the SED of the IR flare to investigate its nature.
The two epochs for the NIR observations were at phases of 567 and 696 days, respectively, around the peak of the MIR flare.
Due to the slow variation in the MIR band, we linearly interpolated the MIR light curves to obtain the MIR fluxes simultaneous with the NIR observations.
The IR SEDs, as shown in Figure~\ref{wisebbfit}(a), can be well fitted with blackbody curves with reduced $\chi^2$ of 2.3 and 3.3 for each epoch.
The blackbody temperatures are $975 \pm 10$ K and $920 \pm 15$ K, respectively.

Assuming that the SED of the IR flare always conforms to blackbody curves, we calculated the blackbody luminosity, temperature, and radius at all MIR epochs, as shown in Figure~\ref{wisebbfit}(b).
The blackbody temperature is $\sim900$ K near the peak, where the luminosity reaches the maximum of $6.0\times10^{43}$ \lum, and drops to $\sim700$ K in the dropping phase.
The blackbody radius is in the range of 0.05 to 0.10 pc.

Integrating the blackbody luminosity with respect to time, we estimated a total IR energy of the flare to be $E_{\rm IR} = (5.1\pm0.1)\times10^{51}$~erg.
This energy was integrated up to Jul 11, 2024, when the flare is still ongoing, suggesting that the above value is only a lower limit.

The SED that matches blackbody curves, the high luminosity of several $10^{43}$~\lum, and the slow variation rule out the synchrotron radiation of a relativistic jet.
Since the blackbody temperature is low with $\sim700-1000$ K, we considered dust thermal emission as the origin of the IR flare.
The energy budget of several $10^{51}$~erg requires the dust to be heated by an energetic transient source, with possible natures being SN, TDE or changing-look AGN (CLAGN).
We will further discuss the possible natures in Section~\ref{Sec 5}.


\subsection{Optical spectra} \label{Sec 3 optical spectra}

\begin{figure*}[!h]
\centering
\begin{minipage}{1.0\textwidth}
\centering{\includegraphics[angle=0,width=1.0\textwidth]{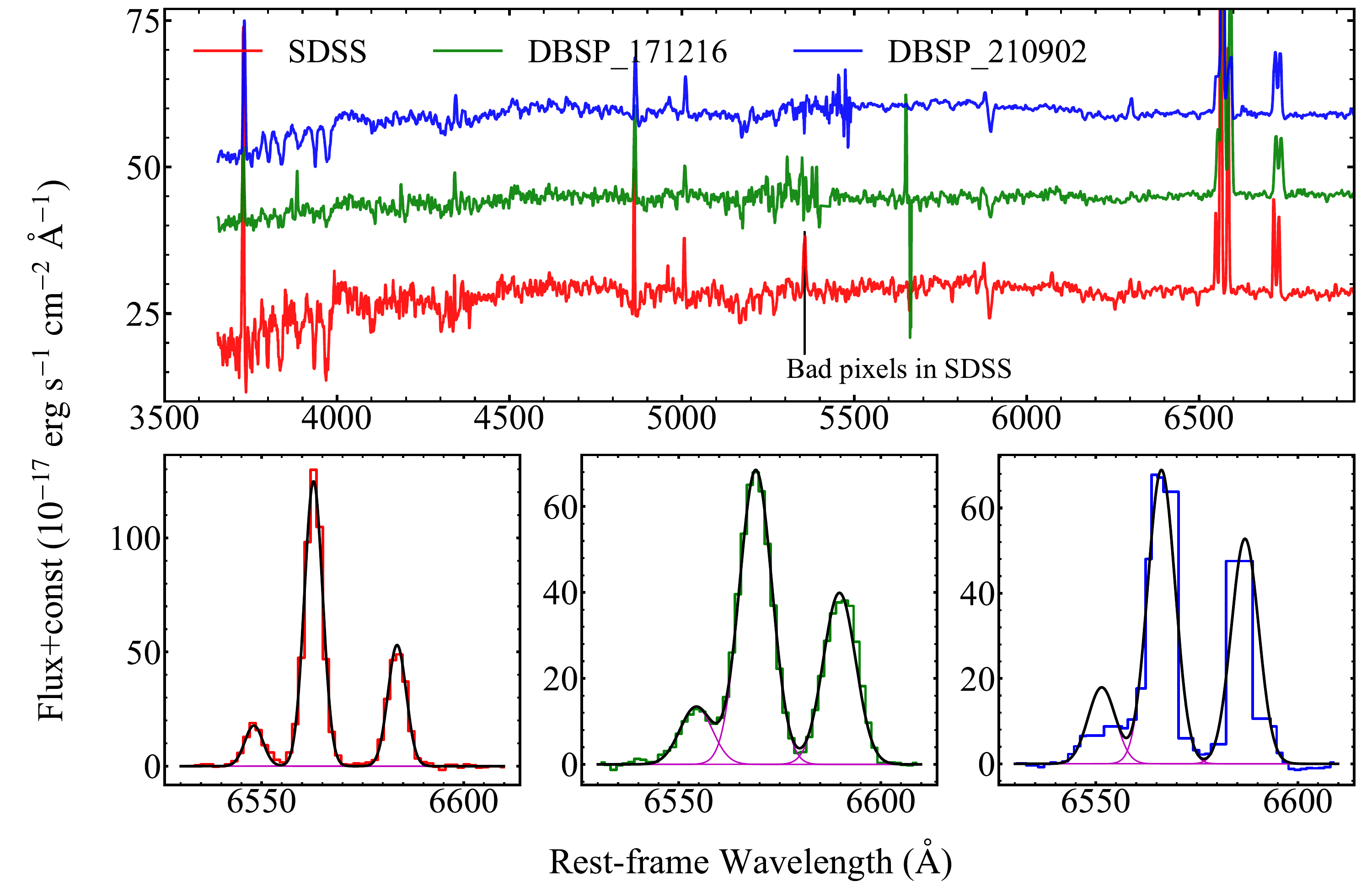}}
\end{minipage}
\caption{Upper Panel: SDSS, DBSP spectra of \objectshortname.
We add a constant to DBSP spectra for clarity.
We labeled the fake feature in the SDSS spectrum due to the bright sky.
Lower panels: the partial enlargement of the starlight-subtracted spectrum around the $\rm H\alpha$ emission lines, and the best-fitting narrow line models.}
\label{spectra}
\end{figure*}

We show the pre-flare SDSS spectrum and the two post-flare DBSP spectra (observed at phases of 355 and 1711 days) in Figure~\ref{spectra}.
The three spectra resemble each other: they are all dominated by starlight with strong narrow emission lines (NEL) detected.
We detected no new emission lines in the post-flare spectra.

We fitted the three spectra with models consisting of a stellar component and NEL components using {\texttt{PPXF}} using a Maximum Penalized Likelihood approach (\citealt{Cappellari2004}).
The model fit the data well for all three spectra with no additional power-law component required.
We present starlight-subtracted spectra in the [N II]+H$\alpha$ region and the best-fitting NEL models in Figure~\ref{spectra}.
No broad emission line component was required.

\begin{figure*}
\centering
\begin{minipage}{1.0\textwidth}
\centering{\includegraphics[angle=0,width=1.0\textwidth]{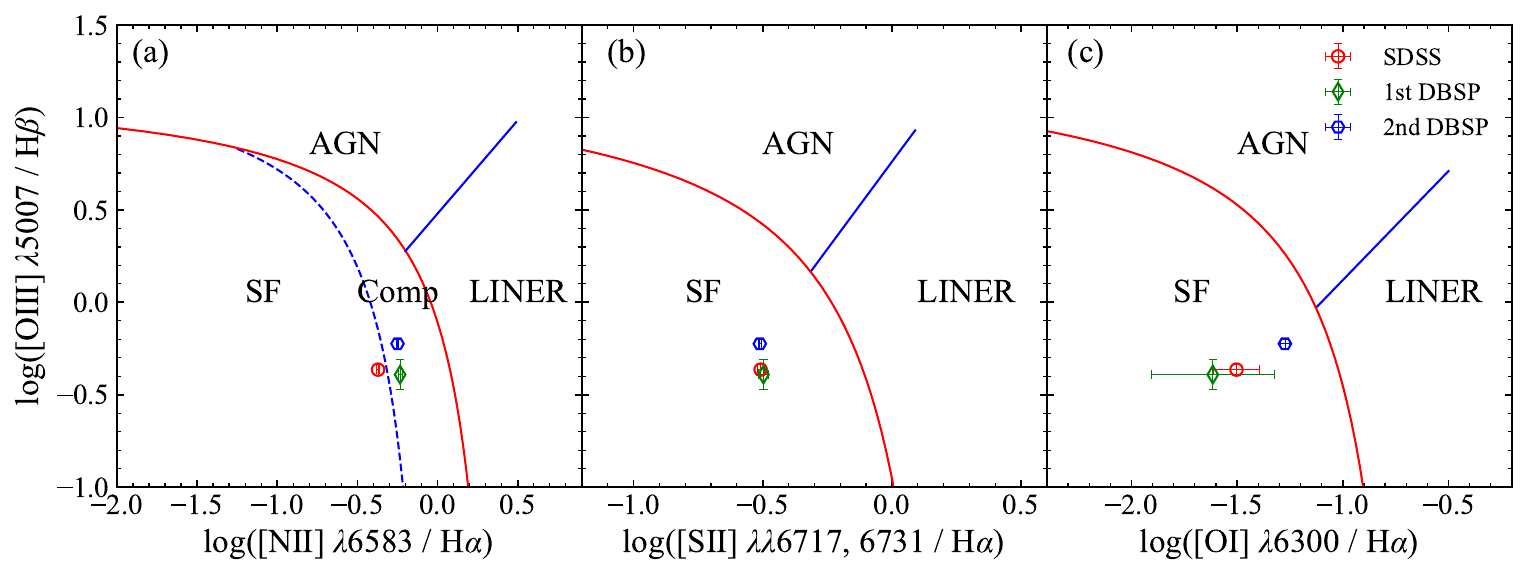}}
\end{minipage}
\caption{BPT diagrams of \objectshortname. \textbf{a,} The [O III]/$\rm H\beta$ verse [N II]/$\rm H\alpha$ diagnostic diagram. The Ke01 (\citealt{Kewley2001}) extreme starburst line and the Ka03 (\citealt{Kauffmann2003}) classification line are shown as the red solid and blue dashed lines, respectively. \textbf{b,} The [O III]/$\rm H\beta$ verse [S II]/$\rm H\alpha$ diagnostic diagram. \textbf{c,} The [O III]/$\rm H\beta$ versus [O I]/$\rm H\alpha$ diagnostic diagram (\citealt{Kewley2006}).}
\label{bpt}
\end{figure*}

We classified the host galaxy of \objectshortname\ using the NEL ratios following \citet{Kewley2001, Kewley2006} and \citet{Kauffmann2003}.
As can be seen in Figure~\ref{bpt}, the NEL ratios on the pre-flare spectrum fall into the star-forming regime, while those on the two post-flare spectra fall into the composite regime due to slightly higher [N II]/H$\alpha$ ratios.
This difference might be caused by the aperture variation of the spectroscopic observations, or by the enhancement of [N II] emission line caused by the IR flare.
In either case, the NELs, especially H$\alpha$ NELs, are mainly excited by star formation in the host galaxy.
Our classification is consistent with that of \citet{Wang2022}, which classified \objectshortname\ as a star-forming galaxy with no broad emission lines based on the SDSS spectrum.

\subsection{Pre-flare SED} \label{Sec 3 pre-flare sed}

We modeled the pre-flare SED with Code Investigating GALaxy Emission (\texttt{CIGALE}; \citealt{Boquien2019}) to measure the properties of the host galaxy.
We assumed a delayed star formation history (SFH) with a single starburst with exponential decay.
We adopted the single stellar population template of \cite{Bruzual2003}, the dust attenuation module based on modified \cite{Calzetti2000} law, the nebular emission model of \cite{Inoue2011}, and the dust emission model of \cite{Dale2014}.
The model fits well with the data with a reduced $\chi^2$ of 0.4, without requiring an additional AGN component.
The inferred stellar mass is $(5.3 \pm 0.7) \times 10^{10}$ \msun\ and the star formation rate is $6.1 \pm 1.3$ \myr.

\begin{figure}[htb]
\centering
\begin{minipage}{0.47\textwidth}
\centering{\includegraphics[angle=0,width=1.0\textwidth]{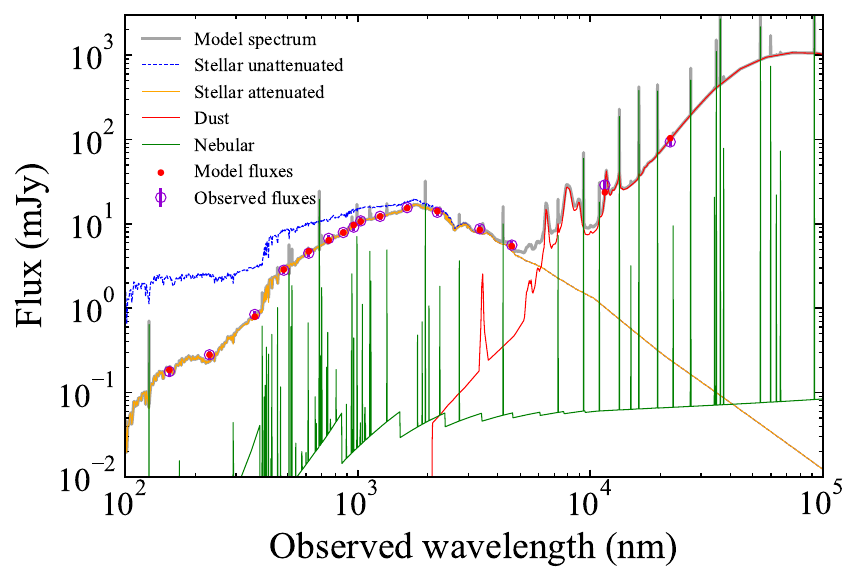}}
\end{minipage}
\caption{The SED of the host galaxy of \objectshortname\ and the best-fitting \texttt{CIGALE} model.}
\end{figure}

We estimated the black hole mass (\mbh) using empirical relations between \mbh\ and the host galaxy properties, including the stellar velocity dispersion (\vdisp) and the bulge mass ($M_{\rm bulge}$) of \citet{KH13} and \citet{McConnell2013}, with typically uncertainties of $0.3 - 0.4$ dex for individual source.
Using the \vdisp\ value of 94~\kms\ measured from the SDSS spectrum with \texttt{PPXF} in section~\ref{Sec 3 optical spectra}, we estimated \mbh\ values of $1.15\times10^7$ and $2.95\times10^6$~\msun, using the relations of \citet{KH13} and \citet{McConnell2013}, respectively.
The difference (0.6 dex) is possibly because the \mbh-\vdisp\ relation at the low mass end (\mbh$\lesssim10^7$~\msun) is less accurate.
\objectshortname\ is a disk galaxy with a bulge accounting for only a small fraction of the total mass.
Using SDSS $r$-band images, \citet{Simard2011} estimated a bulge-to-total flux ratio of 0.14 via bulge-disk structural decomposition.
Combining with the stellar mass from SED fitting, we estimated an $M_{\rm bulge}$ of $7.3\times10^9$~\msun.
The inferred \mbh\ of $1.8\times10^7$ and $2.1\times10^7$~\msun\ using relations from the two literature are in agreement within the typical uncertainty range.
We finally adopted a black hole mass of $\sim1\times10^7$~\msun.

\subsection{No radio variation and radio SED}

Using the measurement results presented in Section~\ref{Sec 2 radio observation}, we did not detect any variation in radio flux in any band.
Thus, a transient radio counterpart associated with the IR flare is not required from the perspective of variability.

\begin{figure}[htb]
\centering
\begin{minipage}{0.45\textwidth}
\centering{\includegraphics[angle=0,width=1.0\textwidth]{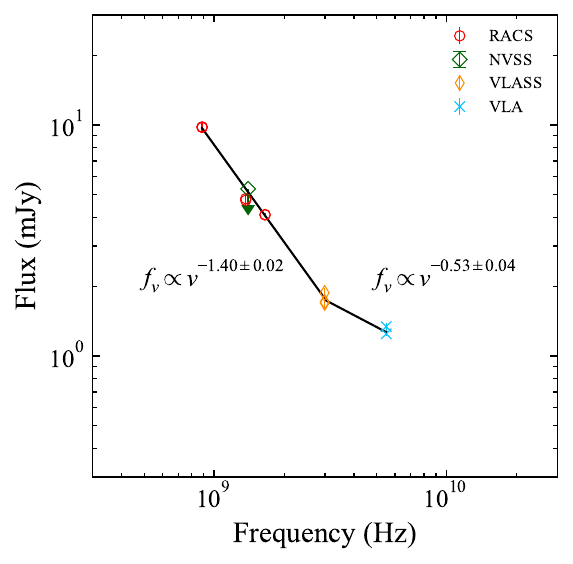}}
\end{minipage}
\caption{$887.5-5500$ MHz radio spectrum of \objectshortname. The solid black lines are the fitted power laws in the $887.5-2988$ MHz and $2988-5500$ MHz bands.}
\label{fig_radio_spectrum}
\end{figure}

Since the host galaxy has a high star formation rate of $\sim6$ \myr, we examined whether the observed radio emission can be explained by star formation.
We present the radio SED in Figure~\ref{fig_radio_spectrum}.
\objectshortname\ shows a steep spectrum with a powerlaw slope $\alpha = 1.40 \pm 0.02$ ($f_{\nu} \propto \nu^{-\alpha}$) below 3 GHz, and a relatively flat spectrum with $\alpha = 0.53 \pm 0.04$ from 3 to 5.5 GHz. 
This SED is consistent with the typical SEDs of star-forming galaxies, which consist of a non-thermal (synchrotron) component dominating at lower frequencies with a mean slope of $\alpha_{\rm nth} \sim 0.6$, and a thermal (free-free) component dominating at higher frequencies with a slope of $\alpha_{\rm th} \sim 0.1$, causing a break in the frequency range of $1-12$ GHz \citep{Klein2018}.
Assuming that the radio flux of \objectshortname\ at 5.5 GHz is thermally dominated, the inferred thermal fraction at 1 GHz is $f_{\rm th, 1\ GHz} \equiv L_{\rm th, 1\ GHz}/L_{\rm 1\ GHz} \sim 0.2$, consistent with the measurement in \cite{Tabatabaei2017} and \cite{Klein2018}.
Using the correlation between the star formation rate and the luminosity at 1.4 GHz \citep{Davies2017}, we estimated a star formation rate of $\sim 5.0$ \myr\ for \objectshortname, consistent with that from SED modeling.

Therefore, we concluded that the radio emission is dominated by host star formation, and there is no evidence of transient radio emission associated with the IR flare.


\section{Dust echo model} \label{Sec 4}

According to the analysis in Section~\ref{Sec 3 sed of the flare and luminosity}, the IR flare in \objectshortname\  originates from dust heated by a transient event with a total energy of $>5\times10^{51}$~erg.
There are two possible ways for dust to be heated.
One is to be heated by the UV/optical radiation of the transient event, known as IR echo \citep{Dwek1983, Graham1983, Miller2010, Jiang2016, vV2016}.
The other is to be heated by the kinetic energy of SN ejecta/TDE outflow released from interaction with circum-stellar/inter-stellar media \citep[e.g.,][]{Fox2011}. 

We estimated the scale of IR emitting dust ($r_{\rm dust}$) using two methods.
First, $r_{\rm dust}$ must be no less than the blackbody radius of 0.05--0.1 pc from the SED analysis.
Second, for the case of IR echo, the primary UV radiation that heats the dust must be more luminous than IR reradiation, and thus we have $L_{\rm UV}>L_{\rm IR}\sim6\times10^{43}$~\lum.
Assuming a sublimation temperature of 1500 K and a typical dust grain radius of 0.1 $\mu$m, the $L_{\rm UV}$ corresponds to a sublimation radius of $\gtrsim0.1$ pc using the relation of \citet{Namekata2016}, and $r_{\rm dust}$ must be even larger.
Both methods yield consistent results that $r_{\rm dust}\gtrsim0.1$ pc.
If the dust is heated by radiation, the IR reradiation would be delayed by $r_{\rm dust}/c\gtrsim0.3$ year, consistent with the observed rise time in the MIR band.
However, if heated by ejecta/outflow with a velocity of $\sim10^4$ km s$^{-1}$, the delay time would be $\gtrsim10$ years.
Such a long time delay makes it difficult to explain the observed sharp rise and rapid decay.
Therefore, we considered dust echo as the origin of the IR flare hereafter.

\begin{figure}[htb]
\centering
\begin{minipage}{0.45\textwidth}
\centering{\includegraphics[angle=0,width=1.0\textwidth]{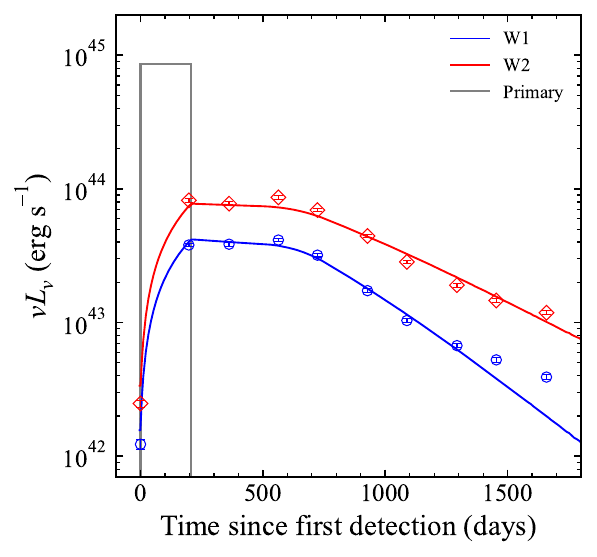}}
\end{minipage}
\caption{The best-fitting IR echo model (requiring the minimum systematic error in the MCMC).
We show the light curve of the primary UV radiation in grey, and the predicted light curves in W1 and W2 bands in blue and red, respectively.
The W2-band data and model are multiplied by 2 for clarity.}
\label{fig_dust_echo}
\end{figure}

\begin{deluxetable}{lr}
\setlength{\tabcolsep}{0.06in}
\tablecaption{Parameters of the dust echo model}
\label{tbl_echo_param}
\tablewidth{0pt}
\tablehead{\colhead{Parameter} & \colhead{Value}}
\startdata
        $t_0$ [MJD]                    & (57742, 57749) \\
        log($L_{\rm 0}$ [erg s$^{-1}$])    & (44.6, 45.0)   \\
        $t_{\rm flare}$ [day]              & (190, 240)  \\
        $r_{\rm in}$ [pc]                  & $<$0.2    \\
        $r_{\rm out}$ [pc]                 & (1.7, 3.2)    \\
        $a_{0}$ [$\rm \mu m$]              & (0.14, 0.25)    \\
        log($n_{\rm d}$ [cm$^{-3}$])         & ($-$9.5, $-$8.5)  \\
        \hline
\enddata
\tablecomments{confidence interval with a 99.7\% probability.}
\end{deluxetable}

To infer the properties of the primary UV radiation, we fitted the observed IR light curves with the dust radiative transfer model of \citet{Lu2016}.
The model was designed to simulate the IR echo of TDEs.
However, since the model can self-consistently calculate the reduction in dust grain radius caused by sublimation through thermal equilibrium, it is also applicable to the study of IR echo from other high-energy transient events, such as SN and changing-look AGN (CLAGN).
The model calculates the IR reradiation of dust in a spherical shell heated by a transient UV source.
The UV source has a luminosity of $L_{\rm 0}$, starting at $t_0$ and lasting for $t_{\rm flare}$ ($t_{\rm TDE}$ in \citealt{Lu2016}).
The dust is a mixture of 47\% graphite and 53\% silicate, the typical value for the Milky Way's dust.
The dust grains with an initial radius of $a_0$ are uniformly distributed in the shell with a number density of $n_{\rm d}$.
The initial inner radius of the shell is $r_{\rm in}$ and the outer radius is $r_{\rm out}$.
We fitted the data using a Markov Chain Monte Carlo (MCMC) approach.
We assumed that all parameters have log-uniform prior probability distributions and that $r_{\rm out}>r_{\rm in}$.
We added a factor to represent the systematic error, whose amplitude is proportional to the model flux.
We only fitted the data for the first 5 years because the later IR emission comes from outer dust, where the assumption of uniformly distributed dust may deviate significantly from the actual situation.

As shown in Figure~\ref{fig_dust_echo}, the model fits well with the data, since it requires a systematic error of only 10\% of the model flux.
This systematic error may be due to the simplicity of our model assumptions.
For example, the UV light curve is oversimplified, the real dust may not be uniformly distributed, or the real dust geometry may not be a spherical shell.
We list the confidence intervals for the parameters in Table~\ref{tbl_echo_param}.
The model predicts that a flare occurred around MJD$\sim57745$, a few days before the first detection in MIR and lasted for $\sim 200$ days.
The luminosity is $\sim(4-10)\times10^{44}$~\lum\ and the inferred total energy is $\sim(0.9-2)\times10^{52}$ erg.

The total energy of the model is higher than that obtained from the SED modeling of $\sim5\times10^{51}$ erg.
We explored the possible causes of this deviation by checking the details of the model.
Using the model parameters, we estimated the radial optical depth of the dust shell as $\tau(\lambda) \approx \pi a_0^2 n_{\rm d} r_{\rm out} Q_{\rm abs}(\lambda) \approx 10 Q_{\rm abs}(\lambda)$, where $Q_{\rm abs}(\lambda)$ is the absorption coefficient.
The inferred UV optical depth $\tau_{\rm UV} \gg 1$, and hence the deviation is not caused by the dust not absorbing all the UV photons.
In Section~\ref{Sec 3 sed of the flare and luminosity}, we found that the spectrum from $J$ to W2 band $1-5\ \mu$m can be described by blackbody spectra.
However, the IR spectrum predicted by the model is a graybody, which mimics a blackbody spectrum at 1--5 $\mu$m where $\tau_{\rm IR}>1$, and is close to emission of optically thin dust at $>5$ $\mu$m where $\tau_{\rm IR}<1$.
In this case, the actual IR luminosity is higher than that calculated by assuming a blackbody spectrum by a factor of a few, thus explaining the deviation.
Note that the graybody spectrum predicted by the model is consistent with the spectroscopic observations of IR TDEs, since their IR emission at $5-15$ $\mu$m is optically thin and shows a strong silicate emission feature \citep{Masterson2025}. 

\section{Discussions} \label{Sec 5}

\subsection{Nature of the IR flare}

\subsubsection{SN?}

Although the host galaxy of \objectshortname\ has a high star formation rate of $\sim5-6$ \myr, predicting a high core-collapse SN rate, the IR flare is unlikely to be caused by a supernova for the following reasons.

First, the position of the flare spatially coincides with the galaxy nucleus, with an offset of $<100$ pc, while most of SNe in spiral galaxies were found off-center.

Second, the IR energy of $\gtrsim5\times10^{51}$ exceeds the total radiant energy of the vast majority of SNe of $\sim10^{51}$ erg.
Only some of super-luminous SNe, a rare type of bright core-collapse SNe, can release so much radiant energy, such as SN 2016aps with $E_{\rm tot} \sim 5 \times 10^{51}$ erg \citep{Nicholl2020}.
However, the host galaxies of super-luminous SNe typically have small masses of $\lesssim10^{9.5}$ \msun\ \citep{Frohmaier2021} and low metallicities \citep{Perley2016}, inconsistent with \objectshortname.

Third, the peak W2 luminosity $L_{\rm W2}$ of $\sim 4\times10^{43}$~\lum\ far exceeds the maximum MIR luminosity of all known SNe.
Based on a sample study of hundreds of Spitzer SNe \citep{Szalai2019}, a check of all SNe IIn exploded during 2009 and 2016 with \wise\ data \citep{Jiang2019}, and a sample study of 11 super-luminous SNe before 2018 with \wise\ data \citep{Sun2022}, the most MIR-luminous SNe have $L_{\rm W2}$ of $\sim(1-2)\times10^{42}$~\lum, an order of magnitude lower than that of \objectshortname.

\subsubsection{CLAGN?}

We found no evidence of AGN in \objectshortname.
The optical spectrum shows neither broad emission lines nor AGN-like narrow emission lines (Section~\ref{Sec 3 optical spectra}).
The pre-flare MIR color of $\rm W1-W2 = 0.18$ is consistent with inactive galaxies and is much bluer than AGNs, which have $\rm W1-W2 > 0.8$ \citep{Stern2012, Yan2013}.
In addition, an AGN component is not required to fit the SED of the host galaxy (Section~\ref{Sec 3 pre-flare sed}).

Although CLAGNs can produce MIR flares, the flares' amplitudes are generally small, with the majority showing $\Delta {\rm W2}<1$ mag and a small fraction showing $\Delta {\rm W2}\approx1.5$ mag \citep{Lyu2022}, where $\Delta {\rm W2}$ is maximum magnitude variation in the W2 band.
Meanwhile, the variation amplitude of \objectshortname\ reaches $\Delta {\rm W2} = 1.9$ mag, exceeding all known CLAGNs in the sample of \citet{Lyu2022}.
In addition, the MIR flares caused by CLAGNs generally rise slowly with rise times of $1-5$ years \citep{Yao2025}, while \objectshortname\ rose in less than half a year.
Therefore, the IR flare in \objectshortname\ is unlikely to be caused by CLAGN.

\subsubsection{Obscured TDE?}

The IR flare in \objectshortname\ rose in several months and remained near the peak for about one year, and these time scales are similar to MIR-bright optical TDEs, such as ATLAS17jrp \citep{Wang2022_17jrp}, ASASSN-18ap \citep{Wang2024}, AT 2020nov \citep{Earl2025}, and AT 2022upj \citep{Newsome2024}, and are also similar to those of WTP14adbjsh discovered in the IR band \citep{Panagiotou2023}.
Its W2-band luminosity of $4\times10^{43}$~\lum\ is slightly higher than MIR bright optical TDEs ($L_{\rm W2}\sim1-2\times10^{43}$~\lum), while is similar to IR TDE Arp 299-B AT1 ($L_{\rm W2}\sim4\times10^{43}$~\lum; \citealt{Mattila2018}).
It shows blackbody spectra in $1-5$ $\mu$m with temperature decreased over time, with $T_{\rm BB}\sim900$ K near the peak and $\sim700$ K in the dropping phase, and the spectra and temperature variation is similar to Arp 299-B AT1 \citep{Mattila2018}.
In short, the IR flare in \objectshortname\ shows many similarities to the IR flares caused by TDEs.

We have inferred the physical parameters of the primary UV flare based on the dust echo model in Section~\ref{Sec 4}.
The peak luminosity of $\sim(4-10)\times10^{44}$~\lum\ and the total energy of $\sim(0.9-2)\times10^{52}$ erg are all within the parameter ranges of known optical TDEs \citep{Lu2018,vV2021, Yao2023}, although in the high end of the samples, possibly because of the selection effect of the MIR sample.
The mass of the central SMBH was estimated to be $\sim 1\times10^7$ $\rm M_\odot$ using the host galaxy properties, not exceeding $10^8$ $\rm M_\odot$ and ensuring that the TDE can occur outside the event horizon.
Therefore, TDE is a reasonable explanation for the origin of the IR flare in \objectshortname.

Assuming a typical temperature of $12000-30000$ K from TDE samples \citep[e.g.,][]{vV2021, Yao2023}, the peak bolometric luminosity inferred from the dust echo model corresponds to a $V$-band peak luminosity of $\sim(2.5-44)\times10^{43}$~\lum.
However, the lack of variation in the binned ASASSN $V$-band light curve limits the variability amplitude to $\lesssim 3.4\times10^{42}$~\lum, indicating a dust extinction of $A_V \gtrsim 2-5$ magnitudes.

\subsection{Obscured TDEs in star-forming galaxies}

Most TDEs' host galaxies are post-starburst \citep[e.g.,][]{French2016,French2020,Law-Smith2017,Hammerstein2021} and few are
star-forming like \objectshortname.
We collected TDEs occurring in star-forming galaxies with ${\rm SFR}>1$ in the literature, including SDSS J095209.56+214313.3 (SDSS J0952+2143 for short, \citealt{Komossa2008,Komossa2009,Palaversa2016}), SDSS J074820.67+471214.3 (SDSS J0748+4712, \citealt{Wang2011}), SDSS J015957.64+003310.5 (SDSS J0159+0033, \citealt{Merloni2015}), IRAS F01004-2237 \citep{Tadhunter2017,Dou2017,Sun2024}, and ASASSN-18ap \citep{Wang2024}.
We list their basic information in Table~\ref{tbl_TDE_SF}.
We found that 4 out of the 5 clearly show luminous IR echoes, while whether SDSS J0159+0033 has an IR echo is unclear due to a lack of IR data.
The high detection rate indicates high dust coverage factors in the nuclei of star-forming galaxies.
This is quite different from TDEs in normal galaxies, which show a weak IR echo with a dust covering factor (defined as the ratio of peak dust luminosity and UV/optical luminosity) of only $\sim1\%$ \citep{Jiang2021b}.

\begin{deluxetable}{lcccr}
\setlength{\tabcolsep}{0.06in}
\tablecaption{Optical/X-ray TDEs in star-forming galaxies}
\label{tbl_TDE_SF}
\tablewidth{0pt}
\tablehead{\colhead{Name} & \colhead{$z$} & \colhead{SFR (\myr)} & \colhead{$t_{\rm peak}$} & \colhead{IR echo}}
\startdata
SDSS J0952+2143  & 0.079 & 1.6  & 2004  & 1,2 \\
SDSS J0748+4712  & 0.062 & 1.0  & <2004 & 2 \\
SDSS J0159+0033  & 0.312 & 5.25 & 2000  & - \\
IRAS F01004-2237 & 0.118 & 120  & 2010\&2021 & 3 \\
ASASSN-18ap      & 0.038 & 5    & 2018  & 4 \\
\enddata
\tablecomments{SFRs of SDSS J0159+0033 and SDSS J0952+2143 were taken from \citet{French2020}, that of SDSS J0748+4712 was taken from \citet{Law-Smith2017}, that of IRAS F01004-2237 was converted from $L_{\rm FIR}=2.6\times10^{45}$ \lum\ \citep{Heckman1987} using relation of \citet{Kennicutt1998}, and that of ASASSN-18ap was taken from \citet{Wang2024}.
The references for the IR echoes are: 1, \citet{Komossa2009}; 2, \citet{Dou2016}; 3, \citet{Dou2017}; 4, \citet{Wang2024}.}
\end{deluxetable}


The IR echo reflects the rich dust at the pc scale in star-forming galaxies.
Combining with the rich dust on a larger scale, which has long been known, the fraction of TDEs being obscured in star-forming galaxies should be much higher than that in other galaxies.
IR-selected samples may directly test this high fraction.
As \citet{Wang2022_17jrp} pointed out, among the 28 MIR flares in star-forming galaxies in the MIRONG sample, only one (ATLAS17jrp) was detected by optical surveys.
In addition, the 5 MIR flares in dusty starburst galaxies in the sample of \citet{Reynolds2022} are all obscured.
The low fraction of optical detection supports the idea that most TDEs in star-forming galaxies are obscured and missed by optical or X-ray surveys, as the case of \objectshortname.

In addition, among the four nearby TDEs detected within 50 Mpc, including IGR J12580+0134 (14.5 Mpc; \citealt{Nikolajuk2013}), WTP14adbjsh (42 Mpc; \citealt{Panagiotou2023}), Arp 299-B AT1 (44.8 Mpc; \citealt{Mattila2018}), and AT2023clx (47.8 Mpc; \citealt{Zhu2023}), two (WTP14adbjsh and Arp 299-B AT1) are obscured TDEs in star-forming galaxies.
This provides another piece of evidence that a significant fraction of TDEs are obscured.

Using 28 MIR flares in star-forming galaxies in the MIRONG sample and the total number of the parent sample, we estimated an event rate of $\sim3\times10^{-5}\ \rm gal^{-1}\ yr^{-1}$ following \citet{Jiang2021a}.
On one hand, although we have ruled out flares in star-forming galaxies with broad emission lines, it is still possible that CLAGN contributes, leading to an overestimation of the rate.
On the other hand, for TDEs in star-forming galaxies, perhaps only those with high UV luminosities, like \objectshortname\ with $L_{\rm UV}\sim(4-10)\times10^{44}$~\lum, can cause MIR variability amplitudes exceeding the threshold of 0.5 magnitude of \citet{Jiang2021a} due to the high MIR background from the host galaxy, leading to an underestimation of the rate.
Despite these uncertainties, the rate is comparable with the observed rate of optical TDEs ($\sim3\times10^{-5}\ \rm gal^{-1}\ yr^{-1}$, \citealt{Yao2023}) and X-ray TDEs ($\sim1\times10^{-5}\ \rm gal^{-1}\ yr^{-1}$, \citealt{Sazonov2021, Grotova2025}), and agrees with the estimation of IR-selected TDE rate $\sim2\times10^{-5}\ \rm {gal}^{-1} \ {yr}^{-1} $ in the IR-luminous galaxy sample of \citealt{Masterson2024}.
This suggests that the actual TDE rate in star-forming galaxies may not be lower than the average of other galaxies if obscured cases are taken into account.

\subsection{Pc-scale dust}

Using the dust radiative transfer model, we inferred a total mass of the echoing dust to be $\sim10-140$ $\rm M_\odot$ within a 3 pc scale.
This mass is $2-3$ orders of magnitude higher than ${\rm log}\ M_{\rm d}=-0.81\pm0.05\ \rm M_\odot$ obtained by \citet{Jiang2021a} by fitting the near-peak SED using an optically-thin dust model.
The reason is that the dust is optically thick at $3-5$ $\mu$m, and that the peak IR emission only comes from a small amount of dust near the paraboloid with the same time delay since the UV illumination does not last long.
Assuming a typical gas-to-dust ratio of 100, the total gas mass in the nucleus reaches $\sim10^3-10^4$ $\rm M_\odot$ within 3 pc.

The dust covering factor at the sub-pc scale of obscured TDEs in \objectshortname\ and analogs can be one to two orders of magnitude higher than that in normal optical TDEs.
The huge difference implies that the sub-pc-scale environment has undergone a drastic evolution from the star-forming to the passive/post-starburst stage.
The transfer may be triggered by intense feedback from star formation coupled with possible nuclear activity, which are both regulated by gas supplies.
Galaxy merging may also play a role by redistributing gas and dust.
Therefore, further efforts of TDE searching by IR echoes can also enable a deeper understanding of the sub-pc environments of SMBHs, which are otherwise extremely challenging to probe.

\section{Summary} \label{Sec 6}

We present the detailed analysis of multi-band data of the IR flare \objectshortname, which was first reported by \citet{Jiang2021a} as a candidate obscured TDE in a star-forming galaxy.
We measured an SFR of $\sim5-6$ \myr\ in the host galaxy based on the SED from UV to MIR and the radio spectrum, and estimated the mass of central SMBH to be $M_{\rm BH}\approx10^7$ \msun\ using the host galaxy's properties.

The IR flare occurred in 2017, peaked in a year, and gradually faded over several years.
It was only detected in the MIR and NIR bands, while no counterpart was detected in the optical, X-ray and radio bands.
The flare's NIR position coincides with the galaxy center, limiting the positional offset to within 100 pc ($3\sigma$ upper limit).
Its $1-5$ $\mu$m SED is consistent with a blackbody curve, whose temperature was $\sim900$ K initially and declined to $\sim700$ K at the late time.
The peak blackbody luminosity is $\sim6\times10^{43}$~\lum\ and the total IR energy is $\sim5\times10^{51}$ erg by integrating the $L_{\rm BB}$ curve.
The IR multi-band light curves can be explained by IR echo, i. e., the IR reradiation from dust heated by a primary UV flare.
We inferred the properties of the primary UV flare using the radiative transfer model of \citet{Lu2016}, and obtained a UV luminosity of $\sim(4-10)\times10^{44}$~\lum, a time duration of $\sim200$ days, and a total UV energy of $\sim(0.9-2)\times10^{52}$ erg.
This total energy exceeds the value estimated by integrating the $L_{\rm BB}$ curve by a factor of $2-4$, because the model predicts that the actual IR SED is a graybody spectrum, which mimics a blackbody spectrum at $1-5$ $\mu$m and exceeds the blackbody's prediction at longer wavelengths. The James Webb Space Telescope (JWST; \citealt{Gardner2006}) working in a wide wavelength range of 0.6 to 29 $\mu$m may provide a powerful way to explore the complete IR SED of SDSSJ0103+1401 and analogs and the characteristics of dust in the future.

We considered three possible origins of the primary UV flare: an SN, a CLAGN or a TDE.
We ruled out the possibility of SN because its position coincides with the center of the galaxy, its total energy exceeds the vast majority of known SNe, and its MIR luminosity exceeds all known SNe by one order of magnitude.
We disfavored the CLAGN explanation because of that no AGN signal was seen before the flare, that the amplitude of the IR flare exceeds the majority of known CLAGNs, and that the rise time is shorter than most known CLAGNs.
TDE is the best explanation because the luminosity, time scale, temperature variation and total energy of the IR flare are consistent with those of known IR bright TDEs.
Since no optical flare was detected, the TDE must be obscured by dust with $A_V\gtrsim2-5$ magnitudes.

This study of \objectshortname\ presents an example of how to identify obscured TDE through IR echo and infer the properties of primary UV radiation.
In the future, similar analyses for more IR flares are required to obtain an unbiased UV luminosity function of TDEs in star-forming galaxies.
In this way, we can ultimately determine whether the star-forming galaxy has a lower TDE rate than post-starburst galaxies, leading to an advanced understanding of the TDE-producing mechanism.

\acknowledgements
We thank the anonymous referee for their very positive and helpful comments, which have improved the manuscript significantly.
This work is supported by the National Key Research and Development Program of China (2023YFA1608100), the National Natural Science Foundation of China (NFSC, grants 12522303,12192221,12393814), and the Strategic Priority Research Program of the Chinese Academy of Sciences (XDB0550200).
L. Sun is supported by NFSC 12103002 and the University Annual Scientific Research Plan of Anhui Province (2022AH010013).
This research uses data obtained through the Telescope Access Program (TAP). Observations obtained with the Hale Telescope at Palomar Observatory were obtained as part of an agreement between the National Astronomical Observatories, Chinese Academy of Sciences, and the California Institute of Technology.

\end{document}